# LIFTED: Household Appliance-level Load Dataset and Data Compression with Lossless Coding considering Precision

Lei Yan, Jiayu Han, Runnan Xu, *Student Member, IEEE*, Zuyi Li, *Senior Member, IEEE*

*Abstract*— The issue of estimating the detailed appliance-level load consumption has received considerable attention. This paper first presents a Labelled hIgh Frequency daTaset for Electricity Disaggregation (LIFTED), which can be used for research on nonintrusive load monitoring (NILM). This dataset consists of one-week detailed appliance-level electricity usage information including voltage, current, active power, and reactive power for a single apartment in the United States, down-sampled at 50Hz. This paper also proposes an efficient Lossless Coding considering Precision (LCP) algorithm on data compression. This algorithm considers both the precision requirements of practical applications on load datasets and the unique characteristics of household appliance-level load datasets. The LCP algorithm is tested on the LIFTED dataset and the results demonstrate that LCP can achieve higher compression ratio compared to several existing algorithms.

## I. Introduction

Recently, sustained growth of electricity consumption has received considerable attention. The incessant demand and tremendous consumption incur concomitant problems, such as the enormous resource and energy consumption as well as severe environmental pollution. In 2018, the electricity consumption of residential consumers accounts for 39% of the total electricity production in the United States, which is the largest proportion of the entire end-use power consumption [1]. An efficient technological solution should be proposed that helps optimize the electricity utilization and provide real-time feedback to residential consumers.

Non-intrusive load monitoring (NILM), first proposed in [2], is the best approach to identifying appliance working states and monitoring appliance power consumption without having direct access to individual appliances. All NILM approaches can be classified into two categories: non-event-based approach and event-based approach. Most of the studies focus on the non-event based approach because of abundant low-resolution sampling dataset such as REDD [3] and AMPds [4]. Event-based approaches take advantage of the transients to help load disaggregation. The BLUED dataset [5] provides labels for each appliance state transition that can be used for event-based approach research, which makes it the first public dataset of its kind to the best of our knowledge. However, this dataset is limited in some applications such as appliance modeling and performance evaluation of load disaggregation due to lack of appliance-level ground truth.

Despite the fact that high-resolution data are rich in informative transients and their relevance to data mining and machine learning techniques can broaden the research fields and improve the accuracy of load disaggregation, there has been relatively little work devoted in this area. The next-generation advanced smart meters and internet of things (IoT) technologies will enable high-resolution load datasets, which, when made publicly available, will promote the development of load disaggregation methods in event-based NILM approaches.

This paper presents the LIFTED dataset, a one-week dataset containing appliance-level details freely available on the web: http://motor.ece.iit.edu/eiot/LIFTED/. This dataset includes residential electricity usage with appliance-level details at 50Hz sampling rate. Like other electricity usage dataset, this dataset could be mined for many applications such as event detection, anomaly detection, appliance fault diagnostics, user behavior analysis, load disaggregation and energy management. Besides, this dataset is specifically geared towards motivating algorithm development of event detection, appliance load modeling for NILM. Since this dataset provides appliance-level ground truth, it could be used to test different NILM algorithms and compare their performance.

In addition to the data collection procedure, this paper also focuses on the compression of big data as well. The large-scale deployment of advanced smart meters will collect massive data so that the data transmission, storage and processing will become an emerging problem. Cloud platform provides virtual storage space and processing capability to help store and further process large volume of data. However, the transmission of large volume of data requires large amount of energy and network bandwidth. So, there is an urgent need of data compression to reduce the size of data. The load data of household appliances can be stored with different precisions for different application purposes using two existing data compression techniques: lossless compression and lossy compression. The proposed data compression algorithm considers the unique characteristics of long steady states and short transients of most household appliance-level load data.

The contributions of this paper are listed as follows.
- To the best of our knowledge, the LIFTED load dataset is the first public high-resolution appliance-level electricity consumption dataset.
- The LCP data compression algorithm can achieve a higher compression ratio for the data series collected from single appliances and smart meters.

The rest of this paper is organized as follows. Section II describes the data acquisition framework for the LIFTED dataset. The LCP data compression algorithm is presented in Section III. Section IV describes the features of the LIFTED dataset, tests the proposed LCP algorithm on the LIFTED dataset, and conducts a NILM test using the LIFTED dataset. Section V concludes this paper and introduces future work.

## II. Data Acquisition Framework

The one-week detailed appliance-level load measurements of the LIFTED dataset were collected from a student apartment in Chicago in September 2019. Hardware and software systems were developed to collect and process the measurements of 15 electrical appliances in the apartment.

*A. Hardware System*

The hardware is a low-cost single-phase device which can be used for high-rate data sampling, efficient calculation, effective filtering, and high-speed data transmission. It mainly includes two modules: signal acquisition and processing (SAP) module that is comprised of current transformer and signal processing integrated circuit (IC), and signal transmission (ST) module containing R485 interface, network interface and center station as shown in Fig. 1.

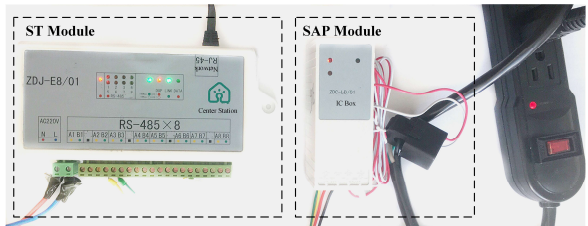

Fig. 1 Picture of data acquisition hardware system

In the SAP module, current measurements are induced through the current transformer made by Zhejiang Shield Electronics CO. which is connected to the power strip. The ADE7953 electrical energy measurement IC made by Analog Devices, Inc. measures line voltage, namely device operating voltage, and current at a high rate up to 6.99KHz, and calculates active, reactive and apparent power, as well as instantaneous rms voltage and current. All the measurements from up to eight different SAP modules are gathered by the center station through RS-485 series interface and sent to the local computer via network interface.

*B. Software System*

After down-sampling at 50Hz, the device number as well as measurements including voltage, current, active power and reactive power are sent to the database on a local computer. As there are 15 appliances in the apartment and the measurement is recorded at 50Hz, about 1,725 MB of data are logged every day, which is economically hard to store in the long run if left uncompressed.

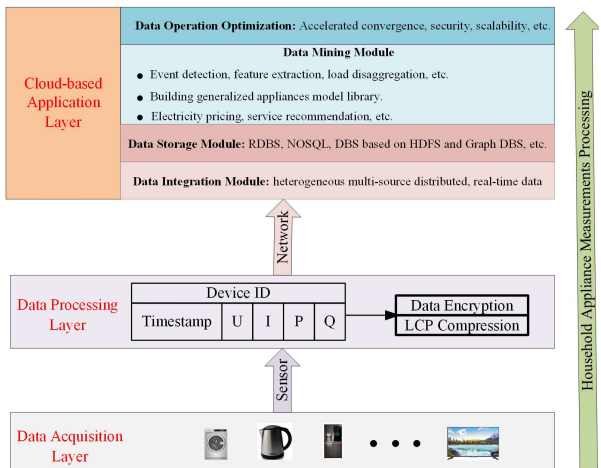

Fig. 2 Framework of household appliances measurements and processing

The software system is designed for receiving and post-processing the raw measurements. The local computer first creates data structure, and then encrypts and compresses the data before transmission. Besides the software running in the local computer, data processing and analysis programs are executed at the cloud computing platform. The programs in the cloud store the measurements with different precisions for different application purposes, build individual appliance load models, generate generalized appliance model library, implement the NILM approaches, and provide electricity usage recommendation service. The framework of data acquisition, storage and processing is described in Fig. 2. In this paper, we mainly introduce the data compression with different precisions.

*C. Privacy Consideration*

With data being transmitted over network and stored in the local computer and cloud platform comes the potential risk for data security and user privacy. In addition, exposing the real time power usage and load disaggregation information on the internet is potentially harmful. For example, it is not difficult to determine whether someone is at home or not based on the detailed appliance-level power usage.

For these reasons, each measurement device is designed with a unique device number. All customers are represented by the device number rather than any other identifying information. Meanwhile, a flexible encoding algorithm is designed to protect the data series, so even if the hackers steal the data, it is hard for them to decrypt the data packet.

### III. DATA COMPRESSION ALGORITHM

Data compression reduces the number of bits needed to represent data which may save storage space significantly. Data compression techniques can be divided into two types: lossless technique and lossy technique. Lossless compression allows the original data to be perfectly reconstructed from the compressed data while lossy compression reconstructs an approximation of the original data. In this paper, a Lossless Coding considering Precision (LCP) data compression algorithm is proposed to compress electrical measurements such as voltage, current, active power and reactive power based on the practical applications in NILM systems.

*A. Data Pattern of Electrical Measurements*

The normal actions of household appliances include turn-on, turn-off, speed adjustment and mode changes which are termed as transients. Following these transients, there are usually a period of times of steady states in which the electrical data do not change significantly compared with the neighbor data points. The duration of transients is generally short whilst that of steady state accounts for nearly the entire running time of appliances as illustrated in Fig. 3.

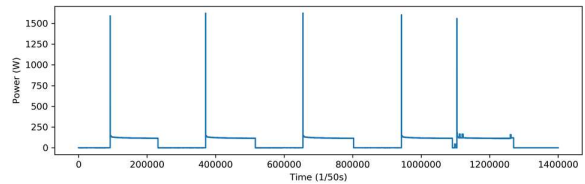

Fig. 3 Cycles of electrical refrigerator.

It should be noted that if any time series data have similar data pattern that the data changes slightly or even remains invariant for a period of times, such data pattern can be taken into consideration when designing efficient data compression algorithm. If the data remains invariant, we need not store the data again or only store the different parts by comparing the current value to the previous one(s).

*B. Data Type and Its Precision*

The LIFTED data acquisition system collects electrical data such as voltage, current, active power and reactive power stored as 32-bit float data format originally. The four columns of data have 2, 3, 2, 2 decimal places, respectively.

As mentioned before, we should take advantage of the characteristic of household appliance load data to design the compression algorithm. Therefore, in order to come up with a more efficient algorithm to achieve this purpose, it is better for the latter data points to be less different with the previous ones. Decreasing the number of decimal places will help make more data points close to the previous ones and achieve higher compression ratio.

Table I Appliance States and Active Power

| Appliance | State | Active Power (W) | Active Power Rounded to Integer (W) |
|---|---|---|---|
| Kettle | 1 | 1027.14 | 1027 |
| Vacuum | 1 | 1001.78 | 1002 |
| Steamer | 1 | 775.38 | 775 |
| Refrigerator | 1 | 41.94 | 42 |
|  | 2 | 123.71 | 124 |
|  | 3 | 165.04 | 165 |
| Washing Machine | 1 | 172.29 | 172 |
|  | 2 | 249.95 | 250 |

Table I shows the states of five common household appliances and corresponding active power. It can be observed that the difference of power consumption for different appliances is significant so that no matter how many decimal places is used to store the data, it will not influence data analysis and relevant applications. Therefore, zero decimal place can be used to format the active power leading to integer type. So is the reactive power.

The application of data varies greatly in different timescales. In short term, the data are mainly used for event detection, anomaly detection, and NILM related applications. However, in long term, the historical data will be used for load forecasting and appliance characteristic analysis which may not need the exact data. All the data can be rounded to numbers ending with '0' and '5' which further decreases the storage space while serving the intended purpose.

For the voltage and current values, keeping two decimal places is enough to conduct related analysis and other applications. As an example, the slight change of appliance operating voltage which is either 120.13V or 120.134V will not affect the quality of data while meeting all the requirements of different applications. So is the current. Therefore, for the household appliance measurements, the voltage, current, power and reactive power take 2, 2, 0, 0 decimal place, respectively. These measurements all become integer types if multiplied by 100, 100, 1, 1, respectively.

*C. Lossless Coding Considering Precision (LCP) Algorithm*

In the computer operating system, all kinds of data such as image, video and audio are stored in binary form. If large abnormal spikes are removed when the data are received, the household load data points would not exceed the range of short int type which is [-32768, 32767]. Under this condition, we can use 16 bits to represent all the measurements and compute XOR of the current and previous values.

The new LCP algorithm proposed in this paper encodes the data series with the following variable length encoding scheme.

1. The first value is stored with no compression.
2. The following eight bits control the number of random values for encryption, i.e., if the eight bits are '00000110', (i.e., 6), six random values will be generated within the range [-32768, 32767]. These values will be encoded together with the collected data.
3. The generated random values in step 2 represent the index where other random values will be inserted. The decompression process will decode and discard these values.
4. In the compression process, if XOR (bitwise exclusive OR) with the previous value is zero, store bit '0'.
5. In the compression process, if XOR with the previous value is non-zero, set three control bits. Store bit '1' on the first space and calculate the number of leading and trailing zeros in the current XOR. Then go to a), b), c) or d):
    a) If both the number of leading zeros and the number of trailing zeros of the current XOR are the same as those of the previous XOR, set the next two control bits to '00' and then store the meaningful bits.
    b) If only the number of leading zeros is the same, set the next two control bits to '01', store the number of trailing zeros in the next 4 bits, and then store the meaningful bits.
    c) If only the number of trailing zeros is the same, set the next two control bits to '10', store the number of leading zeros in the next 4 bits, and then store the meaningful bits.
    d) If neither the number of leading zeros nor trailing zeros is the same, set the next two control bits to '11', store the number of leading zeros in the next 4 bits, store the number of trailing zeros in the next 4 bits, and then store the meaningful bits.

Table II Examples of Integer Values and XOR Operation

| Value | Binary Representation | XOR with Previous |
|---|---|---|
| 23 | 0x0000000000010111 |  |
| 25 | 0x0000000000011001 | 0x0000000000001110 |
| 47 | 0x0000000000101111 | 0x0000000000110110 |
| 48 | 0x0000000000110000 | 0x0000000000011111 |
| 3074 | 0x0000110000000010 | 0x0000110000110010 |
| 3075 | 0x0000110000000011 | 0x0000000000000001 |
| 3076 | 0x0000110000000100 | 0x0000000000000111 |
| 3076 | 0x0000110000000100 | 0x0000000000000000 |

Table III Example of Division in Binary Representation

| Value | Leading Zeros | Meaningful Part | Trailing Zeros |
|---|---|---|---|
| 48 | 0000000000 | 11 | 0000 |

Table IV Example of Lossless Encoding

| Index | Value |
|---|---|
| First XOR | 0x0000000000001110 |
| Control bits | 111 |
| Length of Leading Zeros | 1100 (i.e., 12) |
| Length of Trailing Zeros | 0001 (i.e., 1) |
| Meaningful bits | 111 |

Table II shows examples of binary representation and XOR operation. It can be observed that when the current value is close to the previous value, the number of leading and trailing zeros is nearly the same which need less bits to store the data points based on the proposed encoding method. Table III describes the division of leading zeros, meaningful part, and trailing zeros for the integer 48 as an example. Table IV depicts the encoding method for the first XOR value in Table II. Even though there are three big changes for the eight values in Table II, the total number of bits for storing these values is 103, which is 12.87 bits for a single value and less than the original 16 bits. The number of bits per single value in a practical case is much smaller than the 12.87 in this example, as will be shown in the case study section.

Each floating-point timestamp takes about 4 bytes in the operating system which is quite space consuming. With a fixed data sampling rate, storing only one timestamp in each compressed file is sufficient to reconstruct all timestamps dynamically. This can be helpful to achieve greater compression ratio for sequential electrical data.

It should be noted that the new LCP algorithm proposed in this paper only changes the storage space, rather than changes the numerical magnitude. A similar scheme has been used in the Facebook's time series database (TSDB), namely Gorilla [6]. There are three advantages of the LCP encoding algorithm compared to the Gorilla's compression algorithm:

1. The LCP algorithm takes full advantage of the data type and precision of electrical measurements of household load as described before and represents the measurements by 16 bits other than 64 bits.
2. The LCP algorithm uses a new control bits coding method based on the characteristic of load data to encode the measurements which will improve the compression efficiency greatly.
3. The LCP algorithm generates data packet that is harder to decrypt due to the addition of arbitrary number of values among the data series. If the hacker does not know the encoding rule, they are not likely to decode the data packet. In practical applications, these rules can be easily adjusted.

## IV. Case Study

The LIFTED dataset and the LCP algorithm were presented in the previous two sections. In this section, we summarize the LIFTED dataset and present examples of the LCP algorithm applied to LIFTED. The goal is to explore the potential applications of LIFTED, validate the effectiveness of the LCP algorithm, and present examples of NILM algorithm on LIFTED. All tests run in Python 3.6 on a desktop with a 4.2GHz intel i7-7700K CPU and 16G memory.

### A. LIFTED Dataset

Table I lists part of the appliances that were monitored, together with their working states and corresponding average power consumption. The '0' of OFF state of each appliance was not listed in the table. As shown in Table I, one appliance might have several states with significantly different power and the state index is sorted based on the magnitude of the average power value. The appliances that are not listed in this table are hair dryer, hotpot, mixer, monitor, microwave, rice cooker, toaster, blender, dryer and portable washing machine.

As shown in Fig. 3, the transient process is quite consistent so it can be used as a power signature to identify the state transition from the aggregated data series. Fig. 4 depicts the transition path and parameters of the four different states of the refrigerator in LIFTED. The first value in each path is the transition probability of one appliance running from the starting state to the ending one along the arrow. The second value is the difference between previous steady state average power value and peak value during the transient process, named SPD. The third value is the difference of two steady state average power value, named PPD. It should be noted that the second value is '0' for a power decreasing transition during which there is no peak value. In addition, some appliances do not have steady states such as the computer and TV whose electricity consumptions change time after time. Appliances of this type need to be further studied.

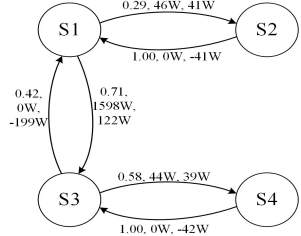

Fig. 4 Transition path and parameters of a refrigerator

LIFTED provides abundant power consumption details such as voltage, current, power and reactive power sampled at 50Hz. The appliances-level information is sufficient to depict steady working states and transitions between different states which can reflect the characteristics of appliances to a certain extent. Appliance-level power consumption details are the cornerstone that promotes the development of researches in the NILM field. Compared with the BLUED dataset, LIFTED provides the ground truth of appliance-level power consumption which could be used for event detection, feature extraction, feature engineering and other relevant research about individual household appliances. All these researches will help build accurate and generalized appliance model which could greatly improve the accuracy of identification. Beyond the researches in NILM field, appliance-level ground truth also helps detect anomaly, diagnose abnormal states. LIFTED complements the existing appliance power consumption datasets which will help extend the research from non-event-based NILM to event-based NILM.

### B. LCP Algorithm Test

In this section we present experimental results for the proposed LCP algorithm and evaluate the performance in terms of compression and decompression speed, compression ratio. Because LCP is a lossless compression algorithm, compression artifact is not evaluated.

The goal of pre-defined data type and its precision is to make the latter data much closer to or even the same as the previous one without affecting normal application. Fig. 5 describes the percentage and average compressed size for each case of the control bits. Roughly 91% of all values are compressed to a single bit since the current and previous values are identical. About 3.92% are compressed with the control bit '111', with an average size of 13.919 bits (or 1.74 bytes). The average size of each value is about 1.768 bits (or 0.221 byte) after compression using the LCP algorithm. Meanwhile the computing speed is about 1,837,982 data per second which is very fast due to the bit operation.

The historical data are more likely to be applied for applications such as load forecasting and demand response which usually do not need the exact values and the approximate values can meet the application requirements. Therefore, we can round it to the nearest integer ending with '0' and '5' at first and then apply LCP to these values which will achieve higher compression ratio. The lossless coding method considering approximation is called LCA.

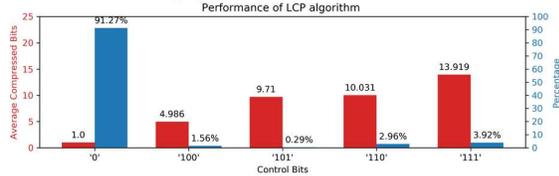

Fig. 5 Distribution of values compressed across different XOR format

Table V compares the compression ratio of the proposed LCP, LCA, Gorilla and Zip algorithms. There are about 15.6 million synthetic data points for the active power of 10.83 hours during which the appliances continuously change their states, and the corresponding file size is 122,228,633 bytes in csv format. In a csv file, one digit is encoded as a byte. For example, a voltage of 120.23 takes 6 bytes.

Table V Comparison of Different Algorithms

| Compression Algorithm | Compressed Size (bytes) | Compression Ratio |
|---|---|---|
| LCA | 2,665,230 | 45.86 |
| LCP | 3,062,652 | 39.90 |
| Gorilla | 21,290,768 | 5.74 |
| Zip | 17,998,745 | 9.22 |

Compared with the data size in csv, the compression ratio (defined as ratio of the size of the original csv file to the size of the compressed file) of LCP is 39.90, which is much larger than that of Gorilla and zip algorithms. The compressed data after LCP are stored in .bin file. The average sizes of data for individual appliances and for their summation is 1.65 bits and 4 bits, respectively. Note that the Gorilla algorithm uses 32 bits (16 bits cannot meet the requirement of conversions between Python values and C structure in Gorilla) to store a single point and the average size for a single compressed data is about 1.5 bytes close to 1.37 bytes in the best case as described in [6].

*C. NILM Test*

In this section we present an example of event-based NILM algorithm applied to LIFTED. The goal of this section is not to explain the NILM algorithm, but rather to demonstrate the good performance of event-based NILM algorithm on LIFTED. The NILM algorithm implements SPD feature to infer states via Maximum A Posteriori (MAP) and conducts steady state verification to further distinguish contributed appliances. Steady state verification is to verify whether the inferred states is correct or not by comparing the estimated total load and the actual total load. Results demonstrates that the SPD feature is so unique that even training with small amount of data with limited number of events could achieve good performance.

To evaluate the performance, we use the synthetic data of ten appliances for test. The first half of individual appliance data are used for training and the summed appliance data of the remaining half are used for testing. The accuracy and precision index defined in (1) and (2) are used to evaluate the performance of the proposed method.

$$Accuracy = \frac{TP + TN}{TP + TN + FP + FN} \quad (1)$$

$$f_1 = \frac{2 \times precision \times recall}{precision + recall} \quad (2)$$

where *TP, TN, FP*, and *FN* represent true positive, true negative, false positive, and false negative, respectively. $Precision = \frac{TP}{TP+FP}$, $Recall = \frac{TP}{TP+FN}$.

The load disaggregation results of 10 selected appliances are shown in Fig. 6. The low $f_1$-score for kettle and vacuum is due to the similarity of their features from ON to OFF leading to false identification. The poor performance of mixer is because its power consumption is so small that some fluctuation in total load will be identified as its turn-off. However, steady state verification is hard to rectify the false identification of small power states during high fluctuation of total load. Thus, feature similarity and identification of small power states are two issues that need to be addressed.

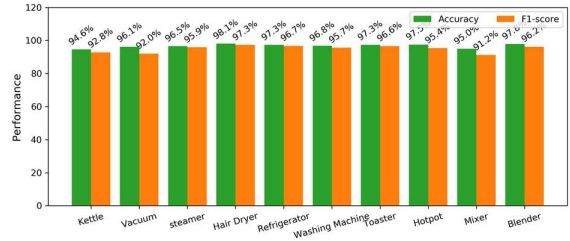

Fig. 6: Performance of load disaggregation on LIFTED

## V. CONCLUSIONS AND FUTURE WORK

This paper introduced LIFTED, a household appliance-level load dataset, and LCP, a data encryption and compression algorithm specially designed for household appliance-level load dataset. The LCP algorithm has been tested on LIFTED and shows good compression performance. An event-based NILM algorithm is also applied on LIFTED and exhibits very accurate appliance identification results.

A future work is to further develop LIFTED that can promote the research towards event-based NILM approaches, where the unique transient features will improve the performance of NILM algorithm greatly. We hope that with the help of LIFTED, more machine learning methods can be used to monitor the appliance states and achieve the goal of reducing electricity consumption.